\documentclass[12pt,preprint]{aastex}

\newcommand{\kms}{km s$^{-1}$}

\slugcomment{}

\shorttitle{Multi-epoch imaging of SiO masers towards TX Cam}
\shortauthors{Diamond and Kemball}

\begin{document}

\title{A movie of a star: multi-epoch VLBA imaging of the SiO masers 
towards the Mira variable TX Cam}

\author{P.J.~Diamond}
\affil{Jodrell Bank Observatory, University of Manchester, 
      Macclesfield, SK11 9DL, UK}
\email{pdiamond@jb.man.ac.uk}

\and
\author{A.J.~Kemball}
\affil {National Radio Astronomy Observatory \\
      P.O. Box 0, Socorro, NM 87801, USA}
\email{akemball@nrao.edu}

\begin{abstract}

We describe an observing campaign using the Very Long Baseline Array
(VLBA) to monitor the time-evolution of the $v=1, J=1-0$ SiO maser
emission towards the Mira variable TX Cam. The data reported here
cover the period 1997 May 24 to 1999 February 19, during which the SiO
maser emission was imaged at approximately bi-weekly intervals.  The
result is an animated movie at an angular resolution of $\sim 500$
$\mu as$, over a full pulsation period, of the gas motions in the near
circumstellar environment of this star, as traced by the SiO maser
emission. This paper serves to release the movie and is the first in a
series concerning the scientific results from this observing
campaign. In this paper, we discuss the global proper motion of the
SiO maser emission as a function of pulsation phase. We measure a
dominant expansion mode between optical phases $\phi \sim 0.7-1.5$
confirming ballistic deceleration, and compare this to predictions
from existing pulsation models for late-type, evolved stars.  Local
infall and outflow motions are superimposed on the dominant expansion
mode, and non-radial local gas motions are also evident for
individual SiO maser components. The overall morphology and evolution
of the SiO emission deviates significantly from spherical symmetry,
with important implications for models of pulsation kinematics in the
near-circumstellar environments of Mira variables.

\end{abstract}


\section{Introduction}

The astronomical SiO masers located in the extended atmospheres of
asymptotic giant branch (AGB) stars act as unique tracers of the
physical processes at work in the near-circumstellar environments of
cool, evolved stars. The extended atmosphere is defined as the region
between the photosphere and the inner dust formation radius. The
individual SiO maser components have sufficiently high brightness
temperatures to allow VLBI radio-interferometric imaging of this
region at sub-milliarcsecond (mas) resolution. The detailed structure
of SiO maser emission in the envelopes of AGB stars has accordingly
been the subject of considerable recent research, much of it
engendered by the imaging capabilities provided by the Very Long
Baseline Array (VLBA) operated by the National Radio Astronomy
Observatory (NRAO\footnote{The National Radio Astronomy Observatory is
a facility of the National Science Foundation operated under
cooperative agreement by Associated Universities, Inc.}). This array
permits routine imaging of 43 GHz SiO maser emission at a resolution
of several hundred $\mu as$. Such observations of the $\nu=1,\ J=1-0$
circumstellar SiO masers towards several late-type evolved stars have
shown that the masers are confined to a narrow, ring-like ellipsoidal
global morphology, and are tangentially amplified \citep{miyoshi94,
diamond94, greenhill95, boboltz97, boboltz00, desmurs00,
cotton03}. The local morphology contains significant fine structure at
radii larger than the inner ring however, including spatially coherent
features and detailed sub-structures.

The VLBA is the first VLBI array to operate continuously throughout
the year and thus offers the possibility of SiO monitoring observations
with both high spatial and temporal resolution. The resulting time
series of high-angular resolution images, when combined as an animated
movie, provides a fundamentally new view of the gas motions in the
near-circumstellar environments of AGB stars.  Animated movies of
astrophysical phenomena are rare despite their unique scientific
value. This is primarily due to the associated technical and
scientific challenges of such observations. Prominent examples of
movies of objects within the solar system include the sequence of
images of comet Shoemaker-Levy 9 colliding with Jupiter
\citep{weaver95} and the complex motions in the solar atmosphere
revealed by the Yohkoh X-ray satellite \citep{aschwanden01}. Examples
of movies of galactic and extra-galactic objects, in both the radio
and the optical, are provided by \citet{tuthill00}, \citet{gomez00},
and \citet{hartigan01}.

We have used the VLBA to create a movie of the 43 GHz, $\nu=1,\ J=1-0$
SiO maser emission towards the Mira variable TX Cam by observing a
closely-sampled time series of high-resolution images. In this paper
we present the first 44 bi-weekly epochs in this series, combined to
produce a movie of the gas motions in the near-circumstellar
environment of TX Cam, as traced by the SiO maser emission during this
period.

Long-period variable (LPV) stars on the AGB, such as TX Cam, have
typical stellar pulsation periods of several hundred days. They have
significant mass-loss during the AGB phase of their evolution, and form
substantial circumstellar shells containing molecular gas and dust. A
full review of this phase of stellar evolution is provided by
\citet{habing96}. The extended atmosphere in AGB stars is a complex
region, dominated by the mass-loss process and permeated by shocks,
magnetic fields and local temperature and density gradients. Models
of the dynamical evolution of the envelopes of evolved stars of this
type require treatment of a range of physical processes, including
stellar pulsation mode, shock propagation, thermal relaxation
mechanisms, dust formation, radiation pressure on dust and models for
the coupling of gas and dust motions, amongst others, and are
generally tractable only through computational approaches
\citep{bowen88,bessell96,willson00,humphreys02}.

A net loss of material occurs from the nearby stellar surface through
the extended atmosphere to the outer stellar wind, although there
remain many uncertainties regarding the physical processes underlying
the mass-loss mechanism. In wind-driven mass-loss models, stellar
pulsation drives shocks into the near-circumstellar environment and
the stellar atmosphere becomes significantly distended relative to a
hydrostatic atmosphere through passage of the outwardly propagating
shocks \citep{bowen88,humphreys02}. This levitates gas above the
photosphere. Subsequently, the gas and circumstellar dust become
coupled and radiation pressure accelerates them outward. In this
mass-loss model, some post-shock material falls back to the stellar
surface during parts of the pulsation cycle, following a trajectory
guided by gravity and local pressure gradients, but there is a net
outflow of material to the outer stellar wind over several pulsation
cycles \citep{bowen88}.

However, there remain many theoretical and empirical uncertainties in
models of the extended atmospheres of AGB stars and the underlying
mass-loss mechanisms. Current theoretical work generally assumes
stationary, polytropic, homogeneous and spherically symmetric
hydrodynamic models \citep[and references therein]{willson00}. Stellar
rotation and magnetic fields are expected to play a significant role
however; supporting evidence is provided by the asymmetric structure
of post-AGB objects and planetary nebulae \citep{garcia99,balick02},
although specific mechanisms for magnetic shaping differ sharply in
interpretation \citep[and references therein]{soker02}.  Direct
observational data are difficult to obtain for the extended
atmospheres around distant cool evolved stars. In this respect, VLBI
offers a unique scientific opportunity to study these objects.

The data presented in this paper are drawn from a long-term VLBA SiO
maser monitoring campaign to compile movies of the gas motions in the
extended atmospheres of a small sample of LPV stars. The overall
scientific goal of this observing program is to constrain models of
these objects by providing empirical data on the detailed kinematics
of the gas and shock motions in the extended atmosphere as well as
information concerning global and local asymmetry. The kinematic and
overall morphological evolution can also be determined as a function
of stellar pulsation phase, which helps to discriminate between
competing physical mass-loss mechanisms. In addition, individual SiO
maser components are usually highly linearly polarized
\citep{mcintosh87,barvainis87,kemball97}, and can be used to probe the
magnitude and orientation of the underlying magnetic field at high
spatial resolution when taken in conjunction with a theoretical model
for the propagation of polarized maser emission. This measurement of
the magnetic field, in turn, constrains gas dynamics in this region. A
final scientific objective of this observing program is to compile
detailed properties of the individual maser components detected in
each epoch, including their size, shape, polarization properties and
proper motions, in order to study component lifetimes, maser
excitation mechanisms and related issues relevant to basic maser
physics.

The variability of SiO masers has been well established through
extensive single-dish observations \citep{martinez88, pijpers94,
alcolea99}.  The net result of these studies is that the integrated
SiO flux appears to vary in phase with the well established optical
variation of Mira variables but with a phase-lag of approximately 20\%
of the pulsation period.  In a recent large-scale statistical
single-dish study, \citet{cho96b} found that the mean velocity of the
v=1 and v=2 SiO masers varied with the optical phase of the star, the
red-shifted emission being dominant between phases 0.3 and 0.8
suggesting that, near optical minimum, the SiO was falling towards the
star. However, this is contradicted by VLBA observations of the proper
motions of v=1, J=1-0 SiO masers in the circumstellar envelope of the
symbiotic Mira R Aqr \citep{boboltz97}. These authors showed that,
between optical phases 0.78 - 1.04 (i.e. just prior to and covering
optical maximum), the masers were falling inwards at a velocity of
approximately 4 km/s. Detailed imaging of an ensemble of several stars
is required to establish the true nature of the mass-loss mechanisms
in AGB stars as a stellar class.

This paper concerns an imaging campaign for the late-type evolved
star, TX Cam, which is a Mira variable with a spectral classification
that varies between M8 and M10 over a pulsation period of 557.4 days
\citep{kholopov85}. The Mira period-luminosity relationship yields a
distance estimate of $\sim 0.39$ kpc for this star \citep{olivier01},
and this distance is adopted here.  SiO maser emission was first
detected by \citet{spencer77}, but there are no associated OH or
H$_2$O masers \citep{dickinson76,wilson72,benson96,lewis97}.
Observations of thermal CO emission lead to an estimated mass-loss
rate of $\sim 1.1 \times 10^{-6} M_{\sun}$/yr \citep{knapp85}. SiO
maser emission has been detected from a range of vibrational and
rotational transitions \citep{jewell87, cho95, bujarrabal96, cho96a,
cho98b, gray99} and SiO isotopomers \citep{barcia89, cho95,
gonzalez96, cho98a}. In addition, several C- and S-rich molecular
species are found in the circumstellar environment \citep{lindqvist88,
bachiller97, olofsson98, bieging00}.

This paper serves to publish the movie produced from the first 44
epochs of the observing campaign, and addresses the global
morphological evolution of TX Cam during this period. Early results
have been reported elsewhere \citep{diamond97,diamond99}. Subsequent
papers will address more detailed aspects of the scientific analysis
of these data. The layout of this paper is as follows. The
observational method and data reduction strategy are briefly discussed
in Section 2. The global properties of the SiO structure and its
evolution are discussed in Section 3 and the summary conclusions are
presented in Section 4.
 
\section{Observations and Data Reduction}

We have observed the 43 GHz $v=1,\ J=1-0$ SiO maser emission towards
TX Cam at multiple epochs using the Very Long Baseline Array
(VLBA). This paper concerns the period 24 May 1997 until 19 Feb 1999,
during which the source was observed approximately bi-weekly, and in
which the optical phase increased from $\phi=0.68 \pm 0.01$ to
$\phi=1.82 \pm 0.01$. The date, project code and optical phase of each
observing epoch are listed in Table 1. The optical phase is computed
using the optical maximum at MJD = 50773 cited by
\citet{gray99}, and assuming their quoted uncertainty $\triangle \phi
\sim 0.01$. A mean period of 557.4 days is adopted \citep{kholopov85}.

Great care was taken to ensure that the same observing schedule was
used for each epoch in order to generate a uniform sequence of
datasets. This included uniformity in both the data acquisition and
correlation configuration. At each epoch the data were recorded in
dual circular polarization in a bandwidth of 4 MHz centered on
$V_{LSR} = 9.0$ km s$^{-1}$, adopting a rest frequency for the $v=1,\
J=1-0$ transition of 43.122027 GHz. Each observing run was of six
hours duration and included interleaved observations of the continuum
calibrators, \objectname{3C454.3}, \objectname{J0359+5057} and
\objectname{J0609-1542}, which were included for residual group delay,
bandpass and polarization calibration. Each schedule was shifted to a
comparable local sidereal time (within a scheduling granularity of one
hour) to minimize a priori differences in $uv-$coverage between the
epochs. Further differences in net $uv$-coverage did arise between
epochs however, due to antenna failures during individual epochs and
also as a result of data editing in post-processing, as described
further below. Representative system temperatures and point source
sensitivities for each 25-meter antenna at 43 GHz were of the order of
$\sim 150$ K and $\sim 11$ Jy/K respectively. The data for each epoch
were correlated in full cross-polarization at the VLBA correlator in
Socorro, adopting a position for \objectname{TX Cam} of
$(\alpha_{J2000}=05^h 00^m 51^s.186\ , \delta_{J2000}=56^d 10^m
54^s.341)$. The correlation accumulation interval was $5 s$, and the
spectra were generated with a sampling of 128 frequency channels over
the 4 MHz band. A restoring beam of 540 $\mu as$ x 420 $\mu as$ at a
P.A. of 20$\degr$ was used in final imaging, representative of the
lowest spatial resolution across the set of observational
epochs. Initial 1024 x 1024 image cubes were generated using uniform
gridding weighting and a pixel spacing of 100 $\mu as$.

The correlated visibility data rate for the project was $\sim 4$ Gbyte
per epoch. Automated pipeline reduction is required as a result of the
data volume alone, but is also dictated by the scientific requirement
for a uniform sequence of images, processed as far as possible through
identical calibration and imaging steps. The data were calibrated and
imaged within the Astronomical Image Processing System (AIPS) using a
POPS script developed by the authors which encodes the data reduction
method described by \citet{kemball95} and \citet{kemball97} for the
reduction of spectral-line VLBI polarimetry observations. The data
reduction heuristics specific to this project were embedded directly
in the scripts; uniform datasets were accordingly essential for the
smooth running of this basic reduction pipeline. The process was
sufficiently automated that the only interaction with the data was
confined to flagging. It was found that careful and ruthless
interactive flagging of the data was essential however in order to
generate images of acceptable quality. The total volume of data
generated in this project ($\sim$ 0.6 Tbytes in all, including all
image cubes and intermediate calibration products) precluded
extracting the ultimate dynamic range from each dataset which might be
possible with highly interactive, custom reduction. Instead, dynamic
ranges of between 100:1 and 200:1 were deemed acceptable.
Occasionally, due to bad weather over parts of the array or the loss
of crucial antennas near the array center in the southwest US,
acceptable images could not be produced. Under these circumstances,
the whole observation was discarded. In the $\sim 80$-week period
covered by these observations, this situation occurred five
times. These epochs are marked in Table 1. In addition, one bi-weekly
observation, near 4 Apr 1998, was not scheduled.

The time sequence of images was combined as an animated movie in order
to explore the overall global evolution of the maser emission in the
circumstellar envelope. A robust velocity moment image was formed from
the Stokes $I$ image cube at each epoch by extracting the maximum
intensity pixel over velocity at each pixel position in projected
right ascension and declination. The animated movie was constructed
from the inner 680 x 680 pixels centered on the projected ring. The
moment images are good tracers of the overall maser structure given
that the masers are tangentially amplified and generally confined to a
narrow projected shell \citep{diamond94}. The singular scientific
advantage in forming an animated sequence of velocity moment images is
that it allows direct visualization and analysis of the gross
morphological evolution of the total SiO maser emission towards
\objectname{TX Cam} in three dimensions rather than four.

Direct astrometry at the $\mu as$ level is not routinely possible for
VLBI observations at 43 GHz. Without phase-referencing, the absolute
astrometric position of each moment image is uncertain. This is the
conventional outcome of VLBI imaging with phase self-calibration, as
used in the reduction of these data. The images were therefore
registered spatially in a separate, two-step process. In the first
pass, the images were aligned based on the assumption that there were
no major changes in overall morphology over the two-week interval
between epochs. The angular offset between successive epochs was then
determined by direct two-dimensional correlation of successive images
using the image registration technique described by \citet{walker97},
and applying the resultant pixel-level shifts to each cube to maximize
the correlation. Cross-correlation image registration techniques are
optimal for aligning identical images but are known to introduce
systematic alignment errors for sufficiently dissimilar images
\citep{brown92}. The TX Cam velocity moment images at each epoch are
sufficiently similar to permit the use of cross-correlation for the
first-order alignment. A feature-based image registration technique
was then used in the second pass to refine the initial frame
offsets. For this method a set of $N_f$ individual component
trajectories $\{(t_i, x_i, y_i);\ i=1,N_j\}_{j=1,N_f}$ were extracted
from the movie for a range of components with continuous motion across
a lower limit of seven or more frames. The centroid of each component
position $(x_i,y_i)$ in the velocity moment image at a given epoch
$t_i$ (frame $k_i$) was measured interactively in the image plane,
with an estimated rms measurement error of 2 pixels (0.2 mas). For
this analysis, no deconvolved component shapes were fitted. These
component trajectories can reasonably assumed to be linear or close to
linear on average. An image registration similarity metric was then
defined as $\sum_{j=1}^{N_f} \chi^2_j$, obtained by fitting a separate
straight line to each measured trajectory. The frame translation
offsets $\{(\triangle x_k, \triangle y_k);\ k=2, N\}$ were then solved
for using Powell's minimization technique. This work was done in the
AIPS++ package using the third-party\footnote{OptSolve++ is
distributed by Tech-X corporation (http://www.techxhome.com)}
optimization library OptSolve++ for an implementation of Powell's
minimization method. In this joint fit for all frame offsets the first
frame was held fixed as the reference point. The second-pass image
registration yielded frame offsets within a range of $\pm 0.6$ mas over
the course of the animated movie, which were then applied as
two-dimensional translations of each frame to yield the final aligned
movie sequence.

For failed and unscheduled observational epochs, as annotated in Table
1 ($\sim 14\%$ of all data), a linearly-interpolated image was
generated for the missing observational epoch and incorporated in the
animated movie to ensure that the interval between frames was
regular. This procedure caused no visible degradation in the
movie. Two image processing filters were applied to each frame to
enhance the visibility of the overall SiO emission in the animated
movie sequence. These included taking the square root of the
brightness $\sqrt{I}$ at each pixel to minimize the range of
amplitudes in the movie. In addition, pixels below $\sqrt{I} < \sim
0.7$ were blanked to suppress emission likely to fall below the
fidelity range of each frame.

The final animated movie can in viewed in MPEG graphics format at
[http//to-be-supplied-by-ApJ].

\section{Results and Discussion}

This section discusses the gross kinematic properties of the SiO maser
emission towards TX Cam in the period covered by these observations.

\subsection{Morphological evolution}

The optical light curve of TX Cam as derived from AAVSO data
(J. Mattei 1998, private communication), is shown in Figure 1, with a
series of markers indicating the times of the VLBA observational
epochs. The VLBA observations, and the associated animated movie
derived from the data, span an optical phase interval of $\phi=0.68
\pm 0.01$ to $\phi=1.82 \pm 0.01$. Individual frames of the movie,
evenly spaced by approximately $\triangle \phi \sim 0.1$, are shown as
intensity contour plots in Figure 2.

In the movie, several key defining properties of the gross morphology
of the SiO maser emission towards TX Cam are clearly visible. Most
predominantly, the maser emission is confined to a narrow projected
ring in all epochs, as found in earlier observations
\citep{diamond94,kemball97}. The projected ring structure is commonly
explained by tangential amplification, arising from a
radial velocity gradient at the inner boundary of the SiO maser region
\citep[and references therein]{humphreys02}. The inner shell boundary
is relatively sharply defined and, in general, has an irregular ellipsoid or
double-ellipsoid structure. Emission which is more diffuse is to be found
outside the inner ring, principally to the east, southeast and north;
such outlier emission is common throughout the period covered by these
observations. There is significant complex fine-structure in the
emission at radii beyond the inner shell, including spatially-coherent
arcs and filaments. There is evidence that some of the outlying
emission features may persist over multiple pulsation cycles. Previous
observations in late 1994 \citep{kemball97} and early 1995 show
complex outlying emission in the east and southeast, similar to the
overall morphology visible in the data presented here.

The clumpiness and granularity of the SiO components in the maser
region reflect a sampling function of the local SiO gas density, maser
pumping conditions and line-of-sight velocity coherence. These are
strongly influenced by anisotropies in the mass-loss process as well
as local turbulent gas motions. Factors of this nature are likely
responsible for the overall E-W asymmetry in the maser emission
intensity evident over large parts of the observing period presented here.

There is significant variability in the structure of the SiO maser
emission over the course of the movie. The gross morphology of the
shell varies smoothly over time, with systematic changes in the
completeness and ellipticity of the inner ring as a function of
stellar phase. The inner ring is well-defined, more complete and more
nearly circular during the initial epochs near an optical phase $\phi
\sim 0.7$. From this phase, through $\phi \sim 1.24$ the overall
structure gradually changes from a well-defined circle to an irregular
ellipsoidal shape with an approximate axis of symmetry at a P.A. of
$\sim 10-30\degr$ (N through E). At about $\phi \sim 1.37$ the SiO
masers in the west and south have faded and are no longer easily
detectable. This condition persists until optical minimum at $\phi
\sim 1.5$, at which point the south-western side of the envelope has
reappeared and a new inner shell boundary starts to form, interior to
the previously expanding shell. The shell becomes progressively better
defined and more complete, bar the northwest quadrant, through the
optical phase $\phi \sim 1.82$ in the closing frame of the
movie. During this final interval the shell takes on an irregular
double-ellipsoid shape, but retains the approximate symmetry axis
noted above. The size of the projected shell at the end of the movie
is comparable to that measured at the starting phase $\phi \sim 0.68$.

The position and evolution of the overall structure and morphology of
the SiO maser emission is not random, in the sense that individual
bright maser features persist over many epochs and show coherent
motions over these periods.  There is also significant variability in
the fine-scale structure of the emission between epochs. Overall, the
strong variability in the spatial structure of the SiO maser emission
shown by these data is entirely consistent with the strong inter- and
intra-cycle variability of SiO spectra known from single-dish
monitoring \citep{martinez88, alcolea99}.

Expansion is the dominant overall kinematic behaviour of the inner
shell boundary between optical phases of $\phi \sim 0.7$ and $\phi
\sim 1.5$. There is significant variability in the fine-scale motions
of individual maser components however. Simultaneous infall is
superimposed on the dominant expansion mode in some parts of the
shell, and during certain phases of the pulsation period. Local
motions, particularly at radii beyond the inner shell boundary, may
show both infall and outflow at the same optical phase. We will
discuss this question in more detail in the following sections. Both
radial and non-radial component motions are present, and non-radial
motions often appear to trace arcs or filamentary structures in the
shell.

The complex spatial structure of the SiO maser emission observed here,
and the high degree of asymmetry have important implications for
theoretical models of the gas kinematics in the near-circumstellar
environment of LPV stars. Standard computational hydrodynamic models
assume a piston-driven pulsation of the photosphere which, in turn,
drives shocks into the extended atmosphere
\citep{bowen88,humphreys96,humphreys02}. These models assume spherical
symmetry in order to reduce the degree of computational complexity and
thus make the models more tractable. The data presented here show
strong evidence for localized mass-loss and a significant departure
from spherical symmetry in the overall morphology and kinematics of
the gas motions in the near-circumstellar environment of TX Cam.

The origin and evolutionary onset of the axisymmetry commonly observed
in many planetary and proto-planetary nebulae remains an important
area of current research \citep{imai02}. It is uncertain when the
asphericity arises in the AGB or post-AGB phase and the dynamical
cause remains unclear. Models have been proposed in which a
circumstellar B-field plays a global dynamical role in shaping the
outflow \citep{garcia99}. Alternative models of localized mass-loss
have also been proposed by \citet[and references therein]{soker02}
based on starspot photospheric cooling over surface regions, leading
to enhanced dust formation, and in which the ratio of magnetic energy
density to thermal energy density in the outflow itself is not
dynamically significant. The influence of binarity, interacting
stellar winds and large-scale photospheric turbulence remain
additional considerations.

The data presented in this paper however, clearly show that spherical
symmetry is a poor assumption for the morphology or kinematics of the
gas motions in the near-circumstellar environments of Mira variables
even during their final evolution on the AGB. We should note however
that while spherical symmetry is clearly a bad approximation for TX
Cam, extrapolating this result to all Miras as a stellar class will
require a larger sample of observed sources covering a range of
evolutionary parameters on the AGB.  We consider this question in
further detail in subsequent papers in this series.

\subsection{Mean shell kinematics}

The TX Cam global shell kinematics can be quantified to first-order by
measuring the variation of mean projected inner shell radius as a
function of observational epoch. This provides a direct measurement of
the mean radial gas motions in the SiO maser zone over the course of
the pulsation cycle. The resulting radial velocity profile can be
compared directly with that predicted by theory and therefore used as
a direct constraint on models of the pulsation hydrodynamics in the
near-circumstellar environment of LPV stars.

Measuring the TX Cam inner shell radius as a function of stellar phase
requires some care given the highly fragmented nature of the inner
shell during large parts of the pulsation cycle. It seldom takes a pure
geometric or mathematical shape and as such is not amenable to direct
and unbiased least-squares fitting in analytic form. In addition, in
such a fit, an objective heuristic would be required to identify the subset
of components which fall within the inner shell boundary and those
which do not. In order to avoid these difficulties, a simple, but more
robust, estimator of the inner shell boundary was developed within the
AIPS++ package, making no assumptions about the geometric form of the
shell itself.  The inner shell radius $r_{in}(\theta)$ at a position
angle $\theta$ on the projected ring, was estimated by finding the
position of the local maximum in the gradient of the radial intensity
function:

\begin{displaymath}
S(r,\theta) = \sum_{\triangle_{(r,\theta)}} I(\eta,\zeta)\ r^{-1}
              d\triangle_{(r,\theta)}
\end{displaymath}

where $I(\eta,\zeta)$ is the velocity moment image brightness at
angular pixel position $(\eta,\zeta)$, and $d\triangle_{(r,\theta)}$
represents summation over a triangular image region with an apex at
the image center pixel at a position angle $\theta$, and of height
$r$. The base of the triangular integration region is therefore
tangential to the projected shell. For each epoch $t_i$, the inner
shell boundary was estimated at each of $N_{\theta}$ position angles
$\theta_m = \frac{2\pi m}{N_{theta}} + \theta_0$, and $r_{in}(t_i)$
was determined as the mean of each $r_{in}(\theta_m)$, weighted by the
magnitude of the radial intensity gradient
$\frac{dS(\theta_m,r)}{dr}_{|r=r_{in}(\theta_m)}$ at that point. The
gradient function at times has multiple peaks at different radii, due
to emission from earlier pulsation cycles. An adaptive windowing
technique was used to maximize the probability of detecting the
current shell in the search for the gradient peak at a given
$\theta_m$. Example results obtained using the inner shell boundary
estimator on the first and last velocity moment frames of the movie,
are shown in Figure 3.

This overall method of estimating $r_{in}(t)$ is inherently
approximate given the fragmented shell structure but does minimize
bias introduced by any assumption of a fixed mathematical shape for
the shell. A quantitative error bar at each epoch was estimated as the
range of $r_{in}$ obtained for each of $N_{\theta} \in \{12,16\}$ and
$\theta_0 \in \{0, \frac{\pi}{N_{\theta}}\}$. The variation in mean
shell radius as a function of stellar phase, computed in this way, is
plotted in Figure 4.

This quantitative measure of the mean shell kinematics as a function
of stellar phase confirms the overall morphological evolution
discussed in the previous section. The expansion between phases $\phi
\sim 0.7$ and $\phi \sim 1.5$ is clearly evident. 
Beyond optical minimum ($\phi > 1.5$) through an optical
phase $\phi \sim 1.8$ in the closing frames of the movie, the
projected shell appears to start contracting as a new inner shell
boundary forms, interior to the previously expanding shell.

The measured radial velocity profile can be compared directly with
model predictions. Theoretical models of the dynamics of LPV
atmospheres have been developed based on numerical studies of
pulsation hydrodynamics in the near-circumstellar environment
\citep{bowen88, bessell96}, as discussed above. Other strong
qualitative constraints on LPV kinematics have also been provided by
optical spectroscopic studies of systematic velocity signatures in
photospheric lines \citep{wallerstein85}. In particular, near-infrared
spectroscopy of the CO $\triangle \nu=3$ vibration-rotation absorption
band at $1.6 \mu m$ has proven an important observational diagnostic
of LPV atmosphere kinematics \citep{hinkle82, hinkle84, wallerstein85,
hinkle97}. The velocity centroid of these lines has a characteristic
S-shaped curve, relative to the velocity center of mass, about a mean
stellar phase of $0.38 \pm 0.05$ \citep{hinkle97}. The velocity curve
is blue-shifted below a phase of $\phi \sim 0.4$, and red-shifted
above, consistent with a shock emerging between pre-maximum and
maximum visible light which rapidly accelerates gas radially outwards
to a velocity of $\sim 20-30$ \kms \citep{hinkle97}. After shock
passage, the gas subsequently decelerates and falls back towards the
photosphere through a stellar phase $\phi \sim 0.8$ under the
influence of gravity and subject to any local pressure gradients
\citep{hinkle82}. The absorption spectra are double-lined near maximum
phase, consistent with simultaneous infall and outflow, above and
below the shock front \citep{alvarez00}. The presence of emission
lines \citep{merrill21} from post-shock material further supports the
shock-wave kinematic model for LPV atmospheres \citep{alvarez00}. 

The $1.6 \mu m$ CO $\triangle \nu=3$ velocity signature traces the
shock kinematics close to the photosphere. For TX Cam, there has been
no direct observational measurement of the photospheric radius
$R_*$. We adopt an estimate $R_*=2.5$ AU, obtained by scaling the
model value reported by \citet{pegourie87} to the adopted TX Cam
distance of $D=0.39$ kpc \citep{olivier01}. The TX Cam SiO maser zone,
centered at $\sim 15.8$ mas, lies at approximately $2.5 R_*$,
assuming these parameters. Numerical models predict the LPV kinematics
at the radius of the SiO maser emission \citep{humphreys02,
bessell96}, although sometimes subject to uncertainties in the
registration of the model and observed stellar phases. The shock
kinematics derived from near-infrared spectroscopy can also be
extrapolated to the larger radius of the SiO maser emission. To
first-order, there is an expected phase offset for the arrival of the
shock in the SiO maser zone due to a propagation delay. From a
ensemble of several Mira variables, \citet{humphreys02} quote a mean
stellar phase of arrival of the shock in the SiO maser zone of $\phi =
0.71 \pm 0.15$. This is consistent with the phase at which a new inner
SiO ring appears in the TX Cam data presented here. A simple
propagation calculation for TX Cam leads to a mean shock front
propagation velocity of $\sim 15$ \kms over this phase
interval. Numerical models by \citet{bowen88} predict a velocity
gradient over the shock fronts located close to the photosphere of
approximately $\sim 20-30$ kms$^{-1}$. Fundamental-mode and
first-overtone pulsators lie at the upper and lower ends of that
velocity range respectively \citep{bowen88}. Velocity amplitudes
derived from near-infrared spectroscopy of a sample of Mira variables
\citep{hinkle97} lie in the range $\sim 23-30$ kms$^{-1}$.

The appearance of a new inner shell in the TX Cam data at a phase
$\phi=1.6 \pm 0.1$ is consistent with an extrapolation of the 1.6 $\mu
m$ shock kinematics to the SiO maser zone at $\sim 2.5 R_*$ and the
mean propagation delay reported by \citet{humphreys02}. From pulsation
hydrodynamic models, the post-shock gas is expected to fall back
ballistically \citep{bowen88}, modified by any local pressure
gradients. The infalling gas will interact with subsequent
outwardly-propagating shock fronts originating in the photosphere.
The outflow and infall time-scales for a local parcel of gas in the
near circumstellar environment vary strongly with radius from the
photosphere however \citep{bowen88,bessell96}. At larger radii, the
reduced gravitational acceleration leads to longer infall times, which
may then exceed the pulsation period. The interaction between these
two time-scales leads to increasing inter-cycle irregularity with
increasing radius, as noted by \citet{bessell96}. For TX Cam at the
adopted mid-point of the SiO maser zone ($\sim 15.8$ mas), the
gravitational acceleration is $g_{SiO} \sim 1.57\ \times 10^{-7}$ km
s$^{-2}$ for a central star mass of 1 $M_{\odot}$, and $g_{SiO} \sim
1.89\ \times 10^{-7}$ km s$^{-2}$ for a central star mass of 1.2
$M_{\odot}$.  For the mean gravitational acceleration, the
corresponding infall time for a parcel of gas across a maser zone of
$\sim 5$ mas width, starting from rest, will exceed the pulsation
period $P=557.4$ days by a factor of $\sim 1-2$. Therefore the
difference in these time-scales can be expected to lead to complex
kinematics in the SiO maser region with an inter-cycle dependence,
dictated by the position at which the infalling and outflowing
material meet and with what relative velocities. In contrast, closer
to the photosphere, the time-scale for ballistic deceleration and gas
infall back towards the photosphere is invariably shorter than the
pulsation period.

To test the hypothesis of ballistic deceleration common in LPV dynamic
models, we fitted a parabolic function to the data presented in Figure
4 over the phases $\phi=0.6$ to $\phi=1.5$. Using the adopted error
bars discussed above for the data points, we derive a mean
deceleration over this region of $1.86\pm 0.26\ \times 10^{-7}$
kms$^{-2}$ and an initial velocity at the inner edge of the SiO maser
region of $7.14 \pm 0.53$ kms$^{-1}$. Therefore our data on TX Cam are
consistent with ballistic deceleration of post-shock SiO gas during
this phase interval of the pulsation period considered here.

\subsection{Component proper motions}

As noted in Section 2, we extracted component trajectories in the
velocity moment images as part of the image registration process. In
Figure 5, we have plotted the projected radial proper motions for four
components from these data, selected for their location on the N, S, E
and W sides of the shell. The radii were computed relative to the mean
shell origin for the first observational epoch, which was held fixed
as a reference in the movie alignment process. A projected radial
velocity was computed at each point by fitting a simple parabolic
model for $r(t)$ in a five-point running window about each data point.

The maser components follow well defined paths but the velocity
magnitudes are not symmetric as a function of position angle on the
shell. This is expected from the global evolution visible in the
animated movie, where considerable non-sphericity is evident over the
pulsation cycle. The radial velocities fall broadly in the range
$5-10$ kms$^{-1}$, consistent with the mean shell kinematics discussed
in the previous section. Limits on the measured variability of radio
continuum flux density from LPV stars at centimeter wavelengths impose
constraints on the possible variability in the temperature and radius
of the radio photosphere (which is defined to lie immediately interior
to the SiO maser zone), and hence on the shock damping which must
apply over radii within $2R_*$ \citep{reid97}. Although individual
maser components shown in Figure 4 have proper motion magnitudes of up
to $\sim 10$ kms$^{-1}$, the measured mean gas motion at the inner
edge of the SiO maser zone of $7.1$ kms$^{-1}$, derived in the
previous section, is comparable to the expected mean upper limit of 5
kms$^{-1}$ derived by \citet{reid97}. In practice, the strong
asymmetry in the near-circumstellar gas motions shown by the data
presented here implies a similar level of asymmetry in the likely
shock damping in the interior radio photosphere.

As discussed above, pulsation hydrodynamic models predict the
simultaneous outflow of gas at inner radii and the infall of gas at
larger radii. The outer gas was levitated above the photosphere in
earlier pulsation cycles. This is quantified in the numerical models
\citep{bowen88} by a predicted saw-tooth pattern in velocity as a
function of radial distance. We do see evidence for this radial
distribution of velocities in the TX Cam data for the pulsation period
presented here. The eastern component in Figure 5 starts beyond the
inner shell radius and falls inwards over part of the cycle, until it
meets the next outwardly propagating material near $\phi \sim
1.6$. Simultaneous infall at larger radii is also evident in the movie
for other individual components, particularly in the north and
south-east parts of the shell. A more detailed proper motion analysis
forms part of a subsequent paper in this series.

\section{Conclusions}

We have monitored the evolution of the 43 GHz $\nu=1,\ J=1-0$ SiO
circumstellar maser emission towards the Mira variable TX Cam at
milliarcsecond spatial resolution over a full pulsation period,
covering a stellar phase range from $\phi=0.68 \pm 0.01$ to $\phi=1.82
\pm 0.01$. This paper presents an animated movie showing the SiO maser
evolution over this period, and discusses the overall kinematic
properties deduced from these data. In summary, we conclude:

\begin{description}

\item{(i) The structure of the SiO maser emission at individual
observational epochs shows morphological features in common with those
found in earlier studies. The maser emission is generally confined to
a narrow, projected ring consistent with tangential amplification. The
inner shell radius is well-defined but more diffuse emission is often
detected at radii beyond the inner boundary. There is significant fine
structure, including coherent arcs and filaments, suggestive of
localized mass-loss. The high degree of spatial variability observed
is consistent with earlier reported single-dish monitoring results.}

\item{(ii) The SiO maser shell shows significant asymmetry, and can
best be described as a fragmented or irregular ellipsoid at many
observational epochs. This non-sphericity has important implications
for current and future theoretical models of pulsation hydrodynamics
in the near-circumstellar environment of LPV stars, which currently
generally assume spherical symmetry.}

\item{(iii) The gas motions, as traced by the SiO emission,
predominantly show expansion over the stellar phase interval
$\phi=0.7$ to $\phi=1.5$. An analysis of the mean shell kinematics
during this interval shows clear evidence for ballistic gravitational
deceleration. Beyond optical minimum at $\phi=1.5$ a new shell forms
interior to the outer shell.}

\item{(iv) Individual maser components have radial proper motion
magnitudes in the range $\sim 5-10$ kms$^{-1}$. The proper motions are
not symmetric as a function of position angle around the SiO maser
shell, and individual maser components may reach up to twice the mean
velocity expected from shock damping limits imposed by known upper
limits for radio continuum variability of LPV stars.}

\end{description}

\acknowledgments

We would like to thank the VLBA scheduling, correlator and operations
staff for making this observing campaign possible. We acknowledge with
thanks, data from the AAVSO International Database, based on
observations submitted to the AAVSO by variable star observers
worldwide. We would like to thank Dr L. Humphreys for her comments on
the manuscript.

\clearpage
\begin{figure}
\figurenum{1}
\epsscale{0.9}
\plotone{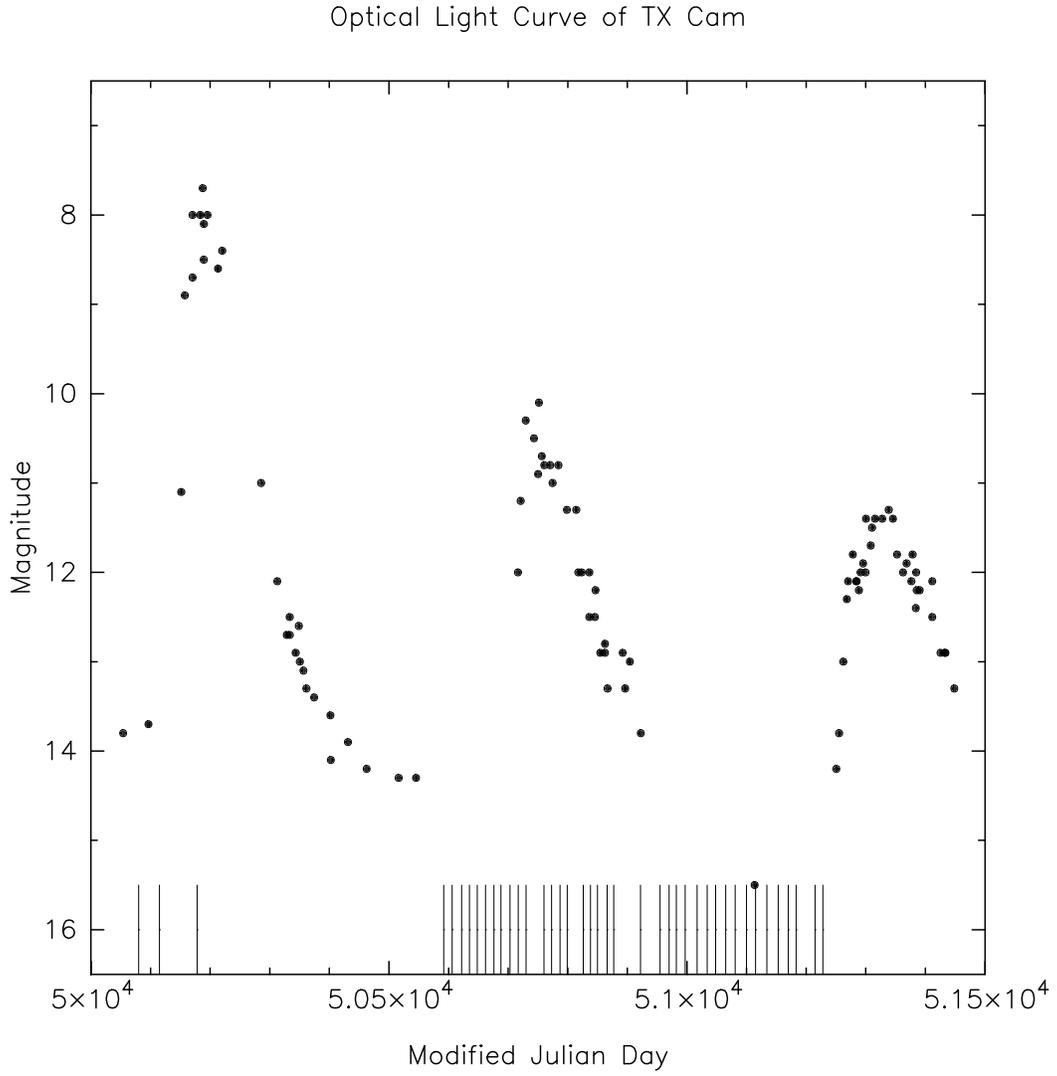}
\caption{The optical light curve of TX Cam as provided by the AAVSO.
The markers at the base of the figure indicate the dates of the VLBA 43 GHz 
observations.}
\end{figure}

\clearpage
\begin{figure}
\figurenum{2}
\epsscale{0.9}
\plotone{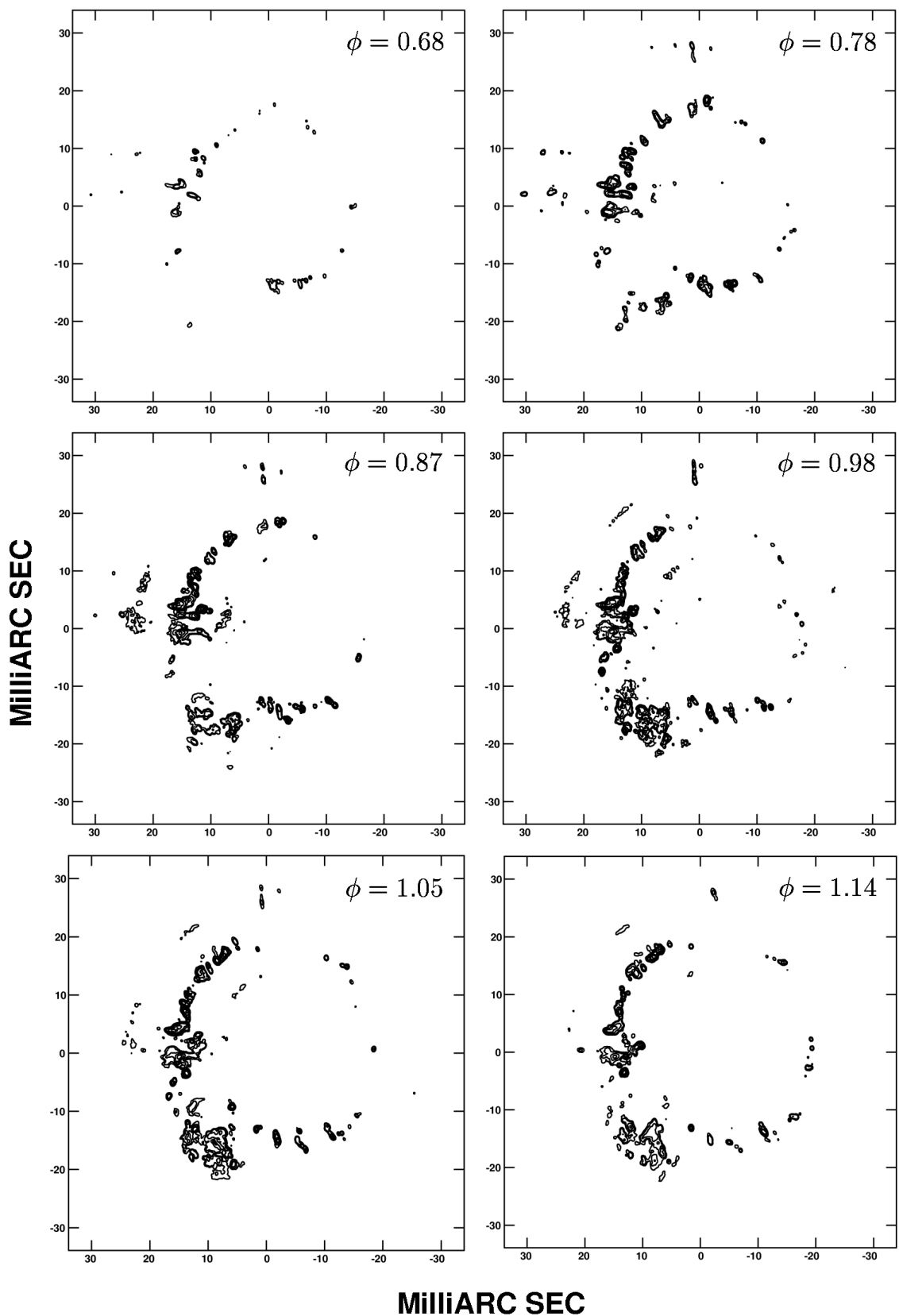}
\caption{A montage of twelve velocity moment images from the TX Cam
movie, spaced at a phase interval of $\triangle \phi \sim 0.1$. Each
frame is displayed as an intensity contour plot with contour levels at
(-10, -5, -2, -1, 1, 2, 5, 10, 20, 40, 80, 100) percent of the frame
peak intensity.}
\end{figure}

\clearpage
\begin{figure}
\figurenum{2}
\epsscale{0.9}
\plotone{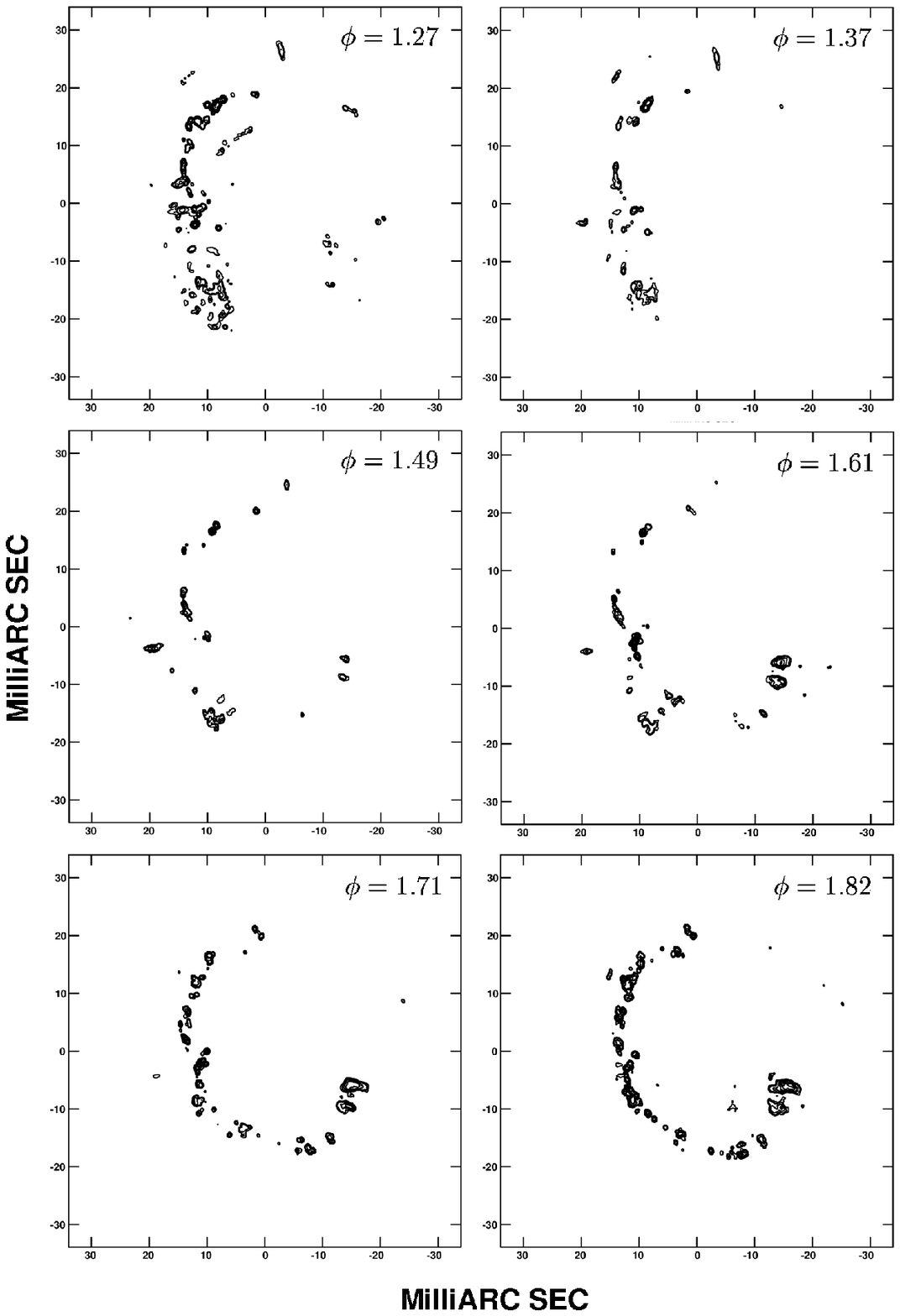}
\caption{Continued.}
\end{figure}

\clearpage
\begin{figure}
\figurenum{3}
\epsscale{0.9}
\plotone{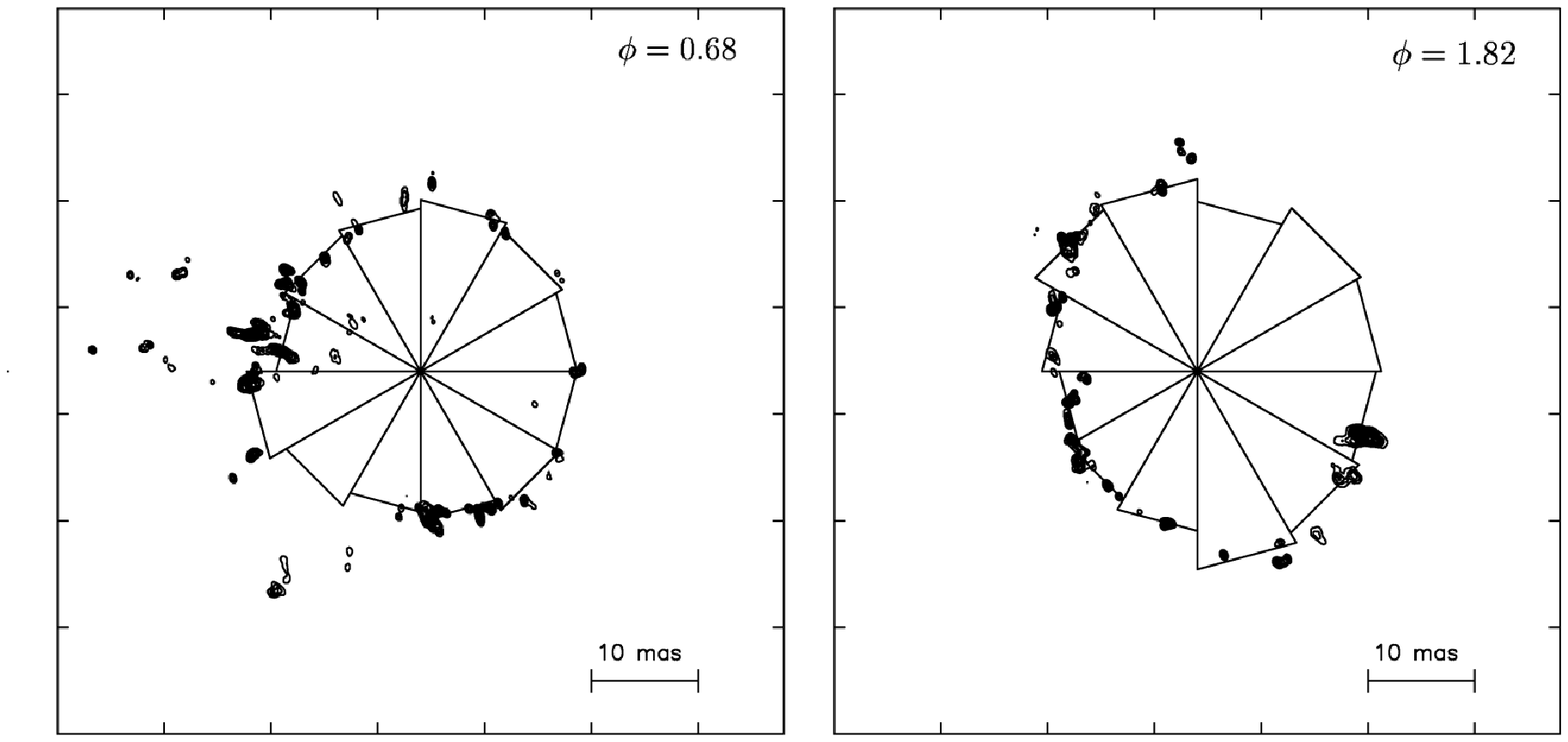}
\caption{Results obtained using the inner shell boundary estimator for
two example velocity moment images, chosen to be the first and last
frames of the movie, at stellar phases of $\phi=0.68$ and $\phi=1.82$
respectively.}
\end{figure}

\clearpage
\begin{figure}
\figurenum{4}
\epsscale{0.9}
\plotone{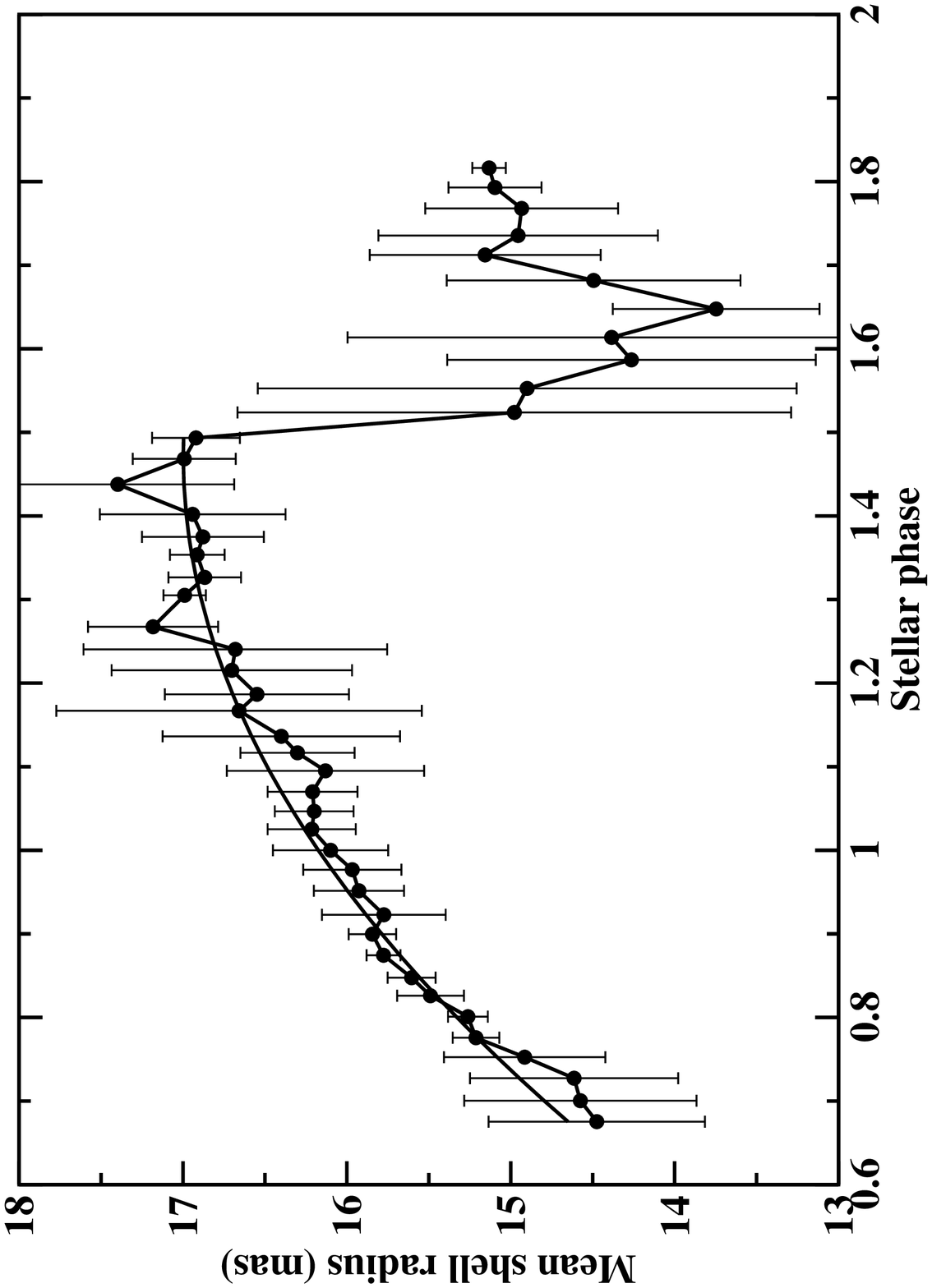}
\caption{Mean inner shell radius as a function of stellar phase.}
\end{figure}

\clearpage
\begin{figure}
\figurenum{5}
\epsscale{0.9}
\plotone{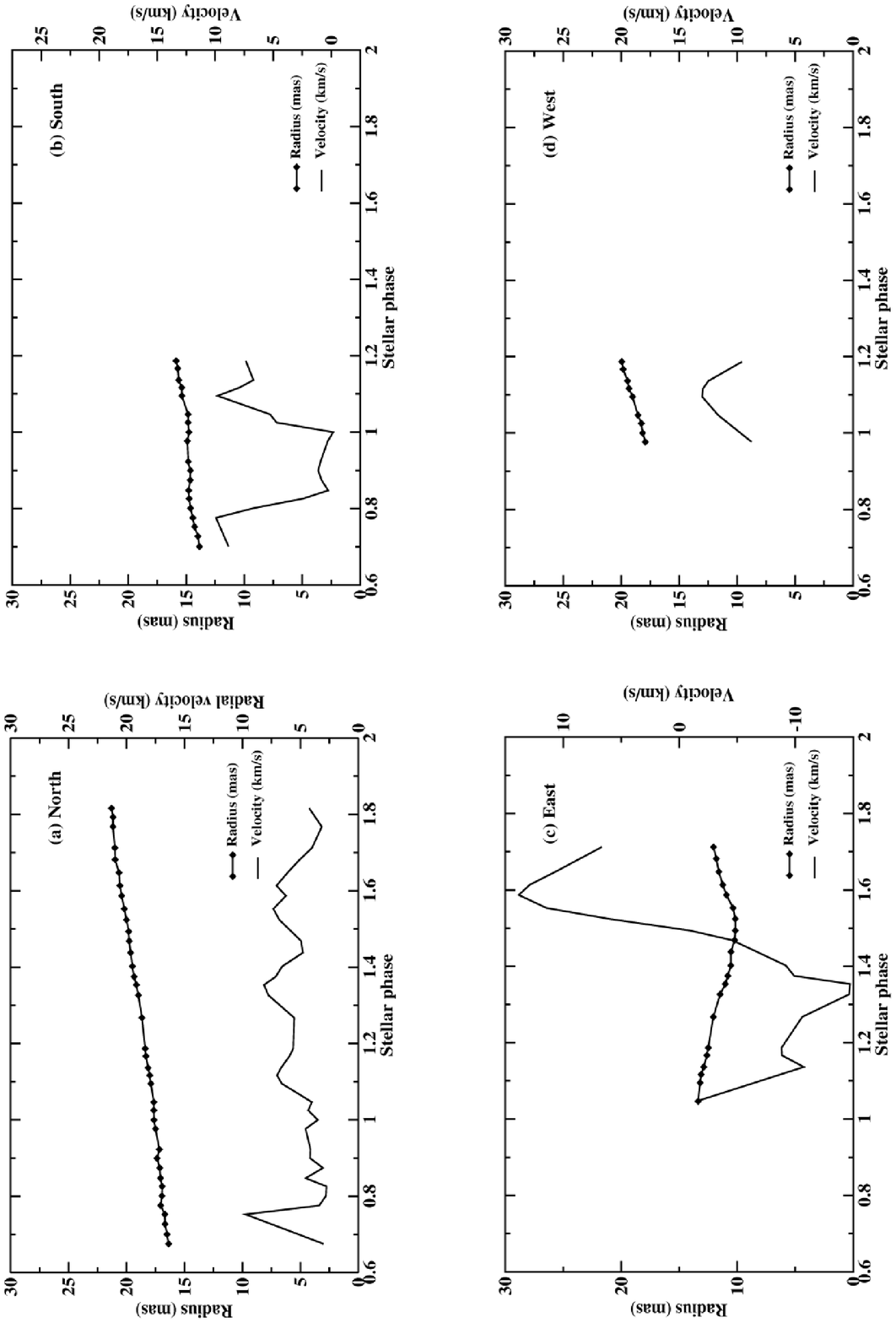}
\caption{Radial proper motions and velocities, as a function of
stellar phase, plotted for four components on the (N,S,E,W) sides of
the projected SiO maser shell.}
\end{figure}

\clearpage
\begin{deluxetable}{llc}
\tablenum{1}
\tablewidth{0pt} 
\tablecaption{Observing dates and epochs}
\tablecolumns{3}
\tablehead{  
\colhead{Epoch code}   &  \colhead{Observing date} & 
\colhead{Optical phase} \tablenotemark{a}\\
 & & $(\phi)$ \\
}
\startdata
BD46A  & 1997 May 24       & 0.68 $\pm$ 0.01 \\
BD46B  & 1997 June 7       & 0.70 $\pm$ 0.01 \\
BD46C  & 1997 June 22      & 0.73 $\pm$ 0.01 \\
BD46D  & 1997 July 6       & 0.75 $\pm$ 0.01 \\
BD46E  & 1997 July 19      & 0.78 $\pm$ 0.01 \\
BD46F  & 1997 August 2     & 0.80 $\pm$ 0.01 \\
BD46G  & 1997 August 16    & 0.83 $\pm$ 0.01 \\
BD46H  & 1997 August 28    & 0.85 $\pm$ 0.01 \\
BD46I  & 1997 September 12 & 0.87 $\pm$ 0.01 \\
BD46J  & 1997 September 26 & 0.90 $\pm$ 0.01 \\
BD46K  & 1997 October 9    & 0.92 $\pm$ 0.01 \\
BD46L \tablenotemark{b}  & 1997 October 25   & 0.95 $\pm$ 0.01 \\
BD46M  & 1997 November 8   & 0.98 $\pm$ 0.01 \\
BD46N  & 1997 November 21  & 0.00 $\pm$ 0.01 \\
BD46O  & 1997 December 5   & 1.03 $\pm$ 0.01 \\
BD46P  & 1997 December 17  & 1.05 $\pm$ 0.01 \\
BD46Q \tablenotemark{b}  & 1997 December 30  & 1.07 $\pm$ 0.01 \\
BD46R  & 1998 January 13   & 1.10 $\pm$ 0.01 \\
BD46S  & 1998 January 25   & 1.12 $\pm$ 0.01 \\
BD46T  & 1998 February 5   & 1.14 $\pm$ 0.01 \\
BD46U  & 1998 February 22  & 1.17 $\pm$ 0.01 \\
BD46V  & 1998 March 5      & 1.19 $\pm$ 0.01 \\
BD46W \tablenotemark{b}  & 1998 March 21     & 1.22 $\pm$ 0.01 \\
\nodata \tablenotemark{c}     & 1998 April 4      & 1.24 $\pm$ 0.01 \\
BD46X  & 1998 April 19     & 1.27 $\pm$ 0.01 \\
BD46Z \tablenotemark{b}  & 1998 May 10       & 1.30 $\pm$ 0.01 \\
BD46AA & 1998 May 22       & 1.33 $\pm$ 0.01 \\
BD46AB & 1998 June 6       & 1.35 $\pm$ 0.01 \\
BD46AC & 1998 June 18      & 1.37 $\pm$ 0.01 \\
BD46AD & 1998 July 3       & 1.40 $\pm$ 0.01 \\
BD46AE & 1998 July 23      & 1.44 $\pm$ 0.01 \\
BD46AF & 1998 August 9     & 1.47 $\pm$ 0.01 \\
BD46AG & 1998 August 23    & 1.49 $\pm$ 0.01 \\
BD46AH & 1998 September 9  & 1.52 $\pm$ 0.01 \\
BD46AI & 1998 September 25 & 1.55 $\pm$ 0.01 \\
BD46AJ & 1998 October 14   & 1.59 $\pm$ 0.01 \\
BD46AK & 1998 October 29   & 1.61 $\pm$ 0.01 \\
BD46AL & 1998 November 17  & 1.65 $\pm$ 0.01 \\
BD46AM & 1998 December 6   & 1.68 $\pm$ 0.01 \\
BD46AN & 1998 December 23  & 1.71 $\pm$ 0.01 \\
BD46AO \tablenotemark{b} & 1999 January 5    & 1.74 $\pm$ 0.01 \\
BD46AP & 1999 January 23   & 1.77 $\pm$ 0.01 \\
BD46AQ & 1999 February 6   & 1.79 $\pm$ 0.01 \\
BD46AR & 1999 February 19  & 1.82 $\pm$ 0.01 \\
\enddata
\tablenotetext{a}{The optical phase is computed using the optical
maximum at MJD = 50773 cited by \citet{gray99}, and assuming their
quoted uncertainty $\triangle \phi \sim 0.01$. A mean period of 557.4
days is adopted \citep{kholopov85}.}

\tablenotetext{b}{Failed epochs, which could not be successfully
reduced in full.}
\tablenotetext{c}{This epoch was not scheduled.}
\end{deluxetable}


\begin{thebibliography}{}

\bibitem[Alcolea et al.(1999)]{alcolea99} 

Alcolea, J., Pardo, J.R., Bujarrabal, V., Bachiller, R., Barcia, A.,
Colomer, F., Gallego, J.D., Gomez-Gonzalez, J., del Pino Cisneros, A.,
Planesas, P., del Rio, S., Rodriguez-Franco, A., del Romero, A.,
Tafalla, M., \& de Vicenta, P. 1999, \aaps, 139, 461

\bibitem[Alvarez et al.(2000)]{alvarez00}
Alvarez, R., Jorissen, A., Plez, B., Gillet, D., \& Fokin, A. 2000,
\aap, 362, 655

\bibitem[Aschwanden, Poland, \& Rabin(2001)]{aschwanden01}
Aschwanden, M.J., Poland, A.I., \& Rabin, D.M. 2001, \araa, 39, 175

\bibitem[Bachiller et al.(1997)]{bachiller97}
Bachiller, R., Fuente, A., Bujarrabal, V., Colomer, F., Loup, C.,
Omont, A., \& de Jong, T. 1997, \aap, 319, 235

\bibitem[Balick \& Frank(2002)]{balick02}
Balick, B., \& Frank, A. 2002, \araa, 40, 439

\bibitem[Barcia, Alcolea, \& Bujarrabal(1989)]{barcia89}
Barcia, A., Alcolea, J., \& Bujarrabal, V. 1989, \aap, 215, L9

\bibitem[Barvainis et al.(1987)]{barvainis87}
Barvainis, R., McIntosh, G., \& Predmore, C.R. 1987, Nature, 329, 613

\bibitem[Benson \& Little-Marenin(1996)]{benson96}
Benson, P.J., \& Little-Marenin, I.R. 1996, \apjs, 106, 579

\bibitem[Bessell, Scholz \& Wood(1996)]{bessell96}
Bessell, M.S., Scholz, M., \& Wood, P.R. 1996, \aap, 307, 481

\bibitem[Bieging, Shaked, \& Gensheimer(2000)]{bieging00}
Bieging, J.H., Shaked, S., \& Gensheimer, P.D. 2000, \apj, 543, 897

\bibitem[Boboltz, Diamond, \& Kemball(1997)]{boboltz97} Boboltz, D.A., Diamond, P.J., \& Kemball, A.J. 1997, \apjl, 487, 147

\bibitem[Boboltz \& Marvel(2000)]{boboltz00} Boboltz, D.A., \& Marvel, K.B. 2000, \apjl, 545, 149

\bibitem[Bowen(1988)]{bowen88}
Bowen, G.H. 1988, \apj, 329, 299

\bibitem[Brown(1992)]{brown92}
Brown, L.G. 1992, ACM Computing Surveys, 24, 325

\bibitem[Bujarrabal et al.(1996)]{bujarrabal96}
Bujarrabal, V., Alcolea, J., S\'anchez Contreras, C., \& Colomer,
F. 1996, \aap, 314, 883


\bibitem[Cho \& Ukita(1995)]{cho95}
Cho, S-H. \& Ukita, N. 1995, \pasj, 47, L1

\bibitem[Cho, Kaifu, \& Ukita(1996a)]{cho96a}
Cho, S-H., Kaifu, N., \& Ukita, N. 1996a, \aaps, 115, 117

\bibitem[Cho, Kaifu, \& Ukita(1996b)]{cho96b}
Cho, S-H., Kaifu, N., \& Ukita, N. 1996b, \aj, 111, 1987

\bibitem[Cho, \& Ukita(1998)]{cho98a}
Cho, S-H., \& Ukita, N. 1998, \aj, 116, 2495

\bibitem[Cho et al.(1998)]{cho98b}
Cho, S-H., Chung, H-S., Kim, H-R., Oh, B-Y., Lee, C-H., \& Han,
S-T. 1998, \apjs, 115, 277

\bibitem[Cotton et al.(2003)]{cotton03}
Cotton, W.D., Mennesson, B., Diamond, P.J., Perrin, G., Cou\'de du
Foresto, V., Chagnon, G., van Langevelde, H.J., Ridgway, S., Waters,
R., Vlemmings, W., Morel, S., Traub, W., Carleton, N., \& Lacasse,
M. 2003, \aap, in press



\bibitem[Desmurs et al.(2000)]{desmurs00}
Desmurs, J.F., Bujarrabal, V., Colomer, F., \& Alcolea, J. 2000, \aap,
360, 189

\bibitem[Diamond et al.(1994)]{diamond94}
Diamond, P.J., Kemball, A.J., Junor, W., Zensus, A., Benson, J., \&
Dhawan, V. 1994, \apj, 430, L61

\bibitem[Diamond, Kemball, \& Boboltz(1997)]{diamond97}
Diamond, P.J., Kemball, A.J., \& Boboltz, D.A. 1997, Vistas Astron,
41, 175

\bibitem[Diamond \& Kemball(1999)]{diamond99}
Diamond, P.J., \& Kemball, A.J. 1999, in {\it IAU 191: Asymptotic
Branch Stars}, eds. T. Le Bertre, A. L\`ebre, \& C. Waelkens
IAU Symposium 191, 195

\bibitem[Dickinson(1976)]{dickinson76}
Dickinson, D.F. 1976, \apjs, 30, 259



\bibitem[Garc\'ia-Segura et al.(1999)]{garcia99}
Garc\'ia-Segura, G., Langer, N., R\'o\.zyczka, M., \& Franco, J. 1999,
\apj, 517, 767

\bibitem[Gomez et al.(2000)]{gomez00}
G\'omez, J.-L., Marscher, A.P., Alberdi, A., Jorstad, S.G., \&
Garc\'ia-Mir\'o, C. 2000, Science, 289, 2317

\bibitem[Gonz\'alez-Alfonso, Alcolea, \& Cernicharo(1996)]{gonzalez96}
Gonz\'alez-Alfonso, E., Alcolea, J., \& Cernicharo, J. 1996, \aap,
313, L13

\bibitem[Gray, Humphreys, \& Yates(1999)]{gray99}
Gray, M.D., Humphreys, E.M.L., \& Yates, J.A. 1999, \mnras, 304, 906

\bibitem[Greenhill et al.(1995)]{greenhill95}
Greenhill, L.J., Colomer, F., Moran, J.M., Backer, D.C., Danchi,
W.C., \& Bester, M. 1995, \apj, 449, 365

\bibitem[Habing (1996)]{habing96}
Habing, H.J. 1996, \araa, 7, 97

\bibitem[Hartigan et al.(2001)]{hartigan01}
Hartigan, P., Morse, J.A., Reipurth, B., Heathcote, S., \& Bally, J. 2001,
\apjl, 559, L157

\bibitem[Hinkle, Hall, \& Ridgway(1982)]{hinkle82}
Hinkle, K.H., Hall, D.N.B., \& Ridgway, S.T. 1982, \apj, 252, 697

\bibitem[Hinkle, Scharlach, \& Hall(1984)]{hinkle84}
Hinkle, K.H., Scharlach, W.W.G., \& Hall, D.N.B. 1984, \apjs, 56, 1

\bibitem[Hinkle, Lebzelter, \& Scharlach(1997)]{hinkle97}
Hinkle, K.H., Lebzelter, T., \& Scharlach, W.W.G 1997, \aj, 114, 2686

\bibitem[Humphreys et al.(1996)]{humphreys96}
Humphreys, E.M.L., Gray, M.D., Yates, J.A., Field, D., Bowen, G., \&
Diamond, P.J. 1996, \mnras, 282, 1359

\bibitem[Humphreys et al.(2002)]{humphreys02}
Humphreys, E.M.L., Gray, M.D., Yates, J.A., Field, D., Bowen, G.H., \&
Diamond, P.J. 2002, \aap, 386, 256

\bibitem[Imai et al.(2002)]{imai02} 
Imai, H., Obara, K., Diamond, P.J., Omodaka, T., \& Sasao, T. 2002,
Nature, 417, 829


\bibitem[Jewell et al.(1987)]{jewell87}
Jewell, P.R., Dickinson, D.F., Snyder, L.E., \& Clemens, D.P. 1987, \apj,
323, 749

\bibitem[Kemball, Diamond \& Cotton(1995)]{kemball95}
Kemball, A.J., Diamond, P.J., \& Cotton, W.D. 1995, \aaps, 110, 383.


\bibitem[Kemball \& Diamond(1997)]{kemball97}
Kemball, A.J., \& Diamond, P.J. 1997, \apj, 481, L111

\bibitem[Kholopov et al.(1985)]{kholopov85}
Kholopov, P.N., Samus, N.N., Frolov, M.S., Goranskij, V.P., Gorynya,
N.A., Kireeva, N.N., Kukarkina, N.P., Kurochin, N.E., Medvedeva, G.I.,
Perova, N.B., \& Shugarov, S.Yu. 1985, General Catalogue of Variable Stars
(Moscow Publishing House:Moscow)

\bibitem[Knapp \& Morris(1985)]{knapp85}
Knapp, G.R., \& Morris, M. 1985, \apj, 292, 640

\bibitem[Lewis(1997)]{lewis97}
Lewis, B.M. 1997, \aj, 114, 1602

\bibitem[Lindqvist et al.(1988)]{lindqvist88}
Lindqvist, M., Nyman, L.-\AA., Olofsson, H., \& Winnberg, A. 1988, \aap,
205, L15

\bibitem[Martinez, Bujarrabal \& Alcolea(1988)]{martinez88}
Martinez, A., Bujarrabal, V., \& Alcolea, J. 1988, AASS, 74, 273

\bibitem[McIntosh(1987)]{mcintosh87}
McIntosh, G.C. 1987, Ph.D thesis, Univ. of Massachusetts.

\bibitem[Merrill(1921)]{merrill21}
Merrill, P.W. 1921, \apj, 53, 185

\bibitem[Miyoshi et al.(1994)]{miyoshi94}
Miyoshi, M., Matsumoto, K., Kameno, S., Takaba, H., \& Iwata, T. 1994, \nat, 
371, 395

\bibitem[Olofsson et al.(1998)]{olofsson98}
Olofsson, H., Lindqvist, M., Nyman, L-\AA., \& Winnberg, A. 1998,
\aap, 329, 1059

\bibitem[Olivier, Whitelock, \& Marang(2001)]{olivier01}
Olivier, E.A., Whitelock, P., \& Marang, F. 2001, \mnras, 326, 490

\bibitem[Pijpers, Pardo, \& Bujarrabal(1994)]{pijpers94}
Pijpers, F.P., Pardo, J.R., \& Bujarrabal, V. 1994, \aap, 286, 501

\bibitem[Pegourie(1987)]{pegourie87}
Pegourie, B. 1987, \apss, 136, 133

\bibitem[Reid \& Menten(1997)]{reid97}
Reid, M.J., \& Menten, K.M. 1997, \apj, 476, 327


\bibitem[Soker (2002)]{soker02}
Soker, N. 2002, \mnras, 336, 826

\bibitem[Spencer et al.(1977)]{spencer77}
Spencer, J.H., Schwartz, P.R., Waak, J.A., \& Bologna, J.M. 1977, \aj,
82, 706

\bibitem[Tuthill et al.(2000)]{tuthill00}
Tuthill, P.G., Monnier, J.D., Danchi, W.C., \& Lopez, B. 2000, \apj,
543, 284

\bibitem[Walker(1997)]{walker97}
Walker, R.C. 1997, \apj, 488, 675

\bibitem[Wallerstein (1985)]{wallerstein85}
Wallerstein, G. 1985, \pasp, 97, 994

\bibitem[Weaver et al.(1995)]{weaver95}
Weaver, H.A., A\'Hearn, M.F., Arpigny, C., Boice, D.C., Feldman, P.D.,
Larson, S.M., Lamy, P., Levy, D.H., Marsden, B.G., Meech, K.J., Noll,
K.S., Scotti, J.V., Sekanina, Z., Shoemaker, C.S., Shoemaker, E.M.,
Smith, T.E., Stern, S.A., Storrs, A.D., Trauger, J.T., Yeomans, D.K.,
\& Zellner, B. 1995, Science, 267, 1282

\bibitem[Willson(2000)]{willson00}
Willson, L.A. 2000, in Unsolved problems in stellar evolution,
ed. M. Livio (Cambridge: Cambridge Univ. Press), 227

\bibitem[Wilson \& Barrett(1972)]{wilson72} 
Wilson, W.J., \& Barrett, A.H. 1972, \aap, 17, 385.

\end{thebibliography}
\end{document}